\documentclass[a4paper, 10pt, twocolumn]{article}

\usepackage{amsfonts,amssymb,amsmath,amsthm}
\usepackage{subfigure}
\usepackage{graphicx}
\usepackage{array}
\usepackage{xcolor}
\usepackage{algorithm}
\usepackage{algorithmic}
\usepackage{cite}
\usepackage[footnotesize]{caption}
\usepackage{tikz}
\usetikzlibrary{positioning}

\usepackage[top=1.5cm, left=1.5cm, right=1.5cm, bottom=1.5cm]{geometry}

\title{Successive Nonnegative Projection Algorithm for Linear Quadratic Mixtures}

\author{Christophe Kervazo$^1$, Nicolas Gillis$^1$ and Nicolas Dobigeon$^{2}$\\
\footnotesize $^1$ University of Mons, Mons, Belgium.\\ \footnotesize $^2$ University of Toulouse, IRIT/INP-ENSEEIHT, 31071 Toulouse Cedex 7, France.\
}
\date{\empty} % no need for a date

\renewenvironment{abstract}{\bf\small {\em\ Abstract---}}{}

\DeclareMathOperator*{\diag}{diag}
\DeclareMathOperator*{\rowrank}{rowrank}

\newtheorem{assumption}{Assumption}[section]

\begin{document}
\newcommand{\norm}[2]{\left\Vert  {#1} \right\Vert _{#2}}
\newcommand{\argmin}[1]{\mbox{argmin}_{#1} \;}
\newcommand{\argmax}[1]{\mbox{argmax}_{#1} \;}
\newcommand*{\Ob}{\mathcal{O}_{b}}
\newcommand{\ckc}[1]{{\color{violet} #1}}
\newcommand*{\ps}{\Pi_\odot}
\newcommand*{\pstr}{\Pi_{\odot^3}}
\newcommand*{\lbr}{[\![ }
\newcommand*{\rbr}{]\!]}
\newcommand*{\snpab}{SNPALQ}
\newcommand{\resi}[2]{\mathcal{R}^f_{#1}\left( #2 \right)}
\newcommand*{\epsb}{\check{\epsilon}}
\newcommand\blfootnote[1]{%
  \begingroup
  \renewcommand\thefootnote{}\footnote{#1}%
  \addtocounter{footnote}{-1}%
  \endgroup
}

% ND's changes
\newcommand{\nd}[1]{\textcolor{red}{#1}}
% ND's remark
\newcommand{\ndr}[1]{\textcolor{red}{(\textbf{ND}: #1)}}

\maketitle
\blfootnote{CK and NG acknowledge the support by the European Research Council (ERC starting grant no 679515), and NG by 
the Fonds de la Recherche Scientifique - FNRS and the Fonds Wetenschappelijk Onderzoek - Vlanderen (FWO) under EOS Project no O005318F-RG47.}
\begin{abstract} 
In this work, we tackle the problem of hyperspectral (HS) unmixing by departing from the usual linear model and focusing on a Linear-Quadratic (LQ) one. The proposed algorithm, referred to as Successive Nonnegative Projection Algorithm for Linear Quadratic mixtures (\snpab), extends the Successive Nonnegative Projection Algorithm (SNPA), designed to address the unmixing problem under a linear model. %relies on the near-separable nonnegative matrix factorization framework. 
By explicitly modeling the product terms inherent to the LQ model along the iterations of the SNPA scheme, the nonlinear contributions in the mixing are mitigated, thus improving the separation quality. The approach is shown to be relevant in a realistic numerical experiment.
\end{abstract}

\section{Introduction}
%\subsection{Linear HS unmixing and near separable non-negative matrix factorization}
HS imaging is a powerful tool in a wide range of fields: remote sensing \cite{Schaepman2009}, biomedical and pharmaceutical imaging \cite{Akbari2011}, astronomy \cite{Themelis2012}, to only name a few. While the datasets are composed of a high number of spectral bands, HS images usually suffer from a limited spatial resolution. Therefore, several materials generally contribute to the measurements associated with each pixel and the acquired spectra correspond to mixtures of pure material spectra, called endmembers.\\
Many works on HS imaging \cite{dobigeon2014nonlinear} have focused on the \emph{linear} mixing model (LMM) which states that the spectral signature of the $i$th observed  pixel $\mathbf{x}_i \in \mathbb{R}^m, i \in \lbr 1,n \rbr$ can be written as
\begin{equation}
\mathbf{x}_i = \sum_{k=1}^{r} h_{ik} \mathbf{w}_k + \mathbf{n}_i
\label{eq:LMM}
\end{equation}
where $\mathbf{w}_k, k \in \lbr 1,r \rbr$, corresponds to the spectral signature of the $k$th endmember, $h_{ik}$ is the the spatial contribution (abundance) of the $k$th endmember in the $i$th pixel and $\mathbf{n}_i$ accounts for any additive noise in the $i$th pixel. In a matrix form, the LMM can thus we rewritten as $\mathbf{X} = \mathbf{W}\mathbf{H} + \mathbf{N}$, with $\mathbf{X} \in \mathbb{R}^{m\times n}$, $\mathbf{W} \in \mathbb{R}^{m\times r}$, $\mathbf{H} \in \mathbb{R}^{r\times n}$ and $\mathbf{N} \in \mathbb{R}^{m\times n}$.\\
Recovering $\mathbf{W}$ and $\mathbf{H}$ from the sole knowledge of $\mathbf{X}$ is referred to as \emph{spectral unmixing} in the HS literature and can be cast as a Blind Source Separation (BSS) problem \cite{comon2010handbook,bobin2015sparsity,kervazo2018PALM}. As the problem is generally ill-posed, additional physical non-negativity constraints are imposed on the unknown matrices $\mathbf{W}$ and $\mathbf{H}$, akin to nonnegative matrix factorization (NMF) \cite{Gillis2014b}. \\ \\

%\subsection{Linear quadratic HS unmixing and NMF}
In various applicative contexts, LMM may however suffer from some limitations and only consists in a $1$st-order approximation. In particular, when the light arriving on the sensor interacts with several materials, nonlinear mixing effects may occur \cite{Bioucas-Dias2012,dobigeon2014nonlinear,Dobigeon2016}. %As such, despite the attractiveness of the linear model stemming from its simplicity, different non-linear models have emerged to tackle cases in which these conditions are not fulfilled \cite{dobigeon2014nonlinear}.\\
To take into account multiple scatterings, bilinear or LQ models include termwise products of the endmembers \cite{Dobigeon2014,Heylen2014}: for all~$i$, 
\begin{equation}
\mathbf{x}_i = \sum_{k=1}^{r} h_{ik} \mathbf{w}_k + \sum_{p=1}^{r}\sum_{l=p+1}^{r} \beta_{ipl} (\mathbf{w}_p\odot \mathbf{w}_l) + \mathbf{n}_i,
\label{eq:LQ}
\end{equation}
where the $\odot$ denotes the Hadamard product and $\beta_{ipl}$ is the contribution of the quadratic term $\mathbf{w}_p\odot \mathbf{w}_l$ in the $i$th pixel.\\
Despite source identifiability issues in the general context of non-linear BSS problem \cite{comon2010handbook,deville2015overview,kervazo2019nonlin}, it was  recently showed that in LQ mixtures the non-linearity leads to a so-called {\it essentially unique} solution\footnote{In the absence of noise and under the additional assumption that $\rowrank(\mathbf{X}) = \frac{r(r+1)}{2}$, if $\hat{\mathbf{W}}$ and $\hat{\mathbf{H}}$ can be found such that $\mathbf{X} = \ps(\hat{\mathbf{W}})\hat{\mathbf{H}}$, then $\hat{\mathbf{W}} = \mathbf{W}$ and $\hat{\mathbf{H}} = \mathbf{H}$ up to a scaling and permutation indeterminacy}, provided that products of the sources up to order four are linearly independent  \cite{Deville2019}. In HS imaging, such an assumption thus requires the family 
\begin{equation}
(\mathbf{w}_i,\mathbf{w}_i\odot \mathbf{w}_j, \mathbf{w}_i \odot \mathbf{w}_j \odot \mathbf{w}_k,\mathbf{w}_i \odot \mathbf{w}_j \odot \mathbf{w}_k \odot \mathbf{w}_l)_{\substack{i,j,k,l \in \lbr 1,r\rbr \\ l < k < j < i}},
\label{eq:linInd}
\end{equation}
whose size scales in $\mathcal{O}(r^4)$, to be linearly independent. This requirement might not be fulfilled in real-world scenario since the number of spectral bands $m$ should then also increase at least as $\mathcal{O}(r^4)$. To overcome this issue, we tackle (\ref{eq:LQ}) under an NMF paradigm. The rationale is to convert the linear independence condition on (\ref{eq:linInd}) into a non-negative independence condition, which is significantly less restrictive in general. 
Specifically, we focus on the so-called Nascimento model defined as \cite{Nascimento2009,Dobigeon2014} 
\begin{equation}
\mathbf{X} = \ps(\mathbf{W}){\mathbf{H}} + \mathbf{N},
\label{eq:LQMat}
\end{equation}
where $\ps(\mathbf{W}) = [\mathbf{w}_i,\mathbf{w}_i\odot\mathbf{w}_j]_{i,j \in \lbr 1,r \rbr, \ j < i} \in \mathbb{R}^{m \times \frac{r(r+1)}{2}}$ is the matrix containing the endmembers $\mathbf{W}$ and their second-order products (referred to as the \textquotedblleft virtual\textquotedblright\ endmembers), and ${\mathbf{H}} \in \mathbb{R}^{\frac{r(r+1)}{2} \times n}$ is the matrix of mixing coefficients associated with the linear and nonlinear contributions, $h_{ik}$ and $\beta_{ipl}$ in (\ref{eq:LQ}), respectively. This model is accompanied by the following constraints
\begin{equation}
\small
\begin{split}
&\forall i \in \lbr 1,n \rbr, \forall k \in \lbr 1,r(r+1)/2 \rbr, h_{ki} \geq 0,\\
&\forall i \in \lbr 1,n \rbr, \sum_{k=1}^{\frac{r(r+1)}{2}} h_{ki} \leq 1,\\
%&\forall i\in \lbr 1,m \rbr, \forall k \in \lbr 1,r \rbr, 0 \leq \mathbf{W}_i^k \leq 1.
%&\alpha = \min_{\substack{j\in \lbr 1,r \rbr \\ x\in \Delta}} \norm{\mathbf{W}_j - \ps(\mathbf{W})_\mathcal{J}x}{2} > 0,\ \mathcal{J} = \lbr 1,\tilde{r} \rbr \setminus \{j\}
\end{split}
\label{eq:hypLQEntiere}
\end{equation}
Concerning $\mathbf{W}$, no endmember must lie within the convex hull formed by the other (virtual) ones and the origin. Lastly, the mixing is assumed to be LQ near-separable, which generalizes the pure pixel assumption \cite{Gillis_12_FastandRobust,Ma2013}:
\begin{assumption}
	$\mathbf{X}$ is said $r-$LQ near-separable\footnote{Note that the virtual endmembers are not required to appear as pure pixels, prohibiting the mere use of linear near-separable NMF algorithms.} if it can be written as:
	%	\begin{equation*}
	%	\mathbf{X} = \ps(\mathbf{W})\begin{bmatrix}
	%	\mathbf{I_r} & \mathbf{H'}   \\
	%	\mathbf{0}_{\frac{r(r-1)}{2} \times r} &   
	%	\end{bmatrix}\mathbf{\Pi} + \mathbf{N},
	%	\end{equation*}
	\begin{equation}
	\small
	\mathbf{X} = \ps(\mathbf{W})
	\underbrace{
	\left[\begin{array}{cc}
	\begin{array}{c}
	\mathbf{I_r}\\
	\mathbf{0}_{\frac{r(r-1)}{2} \times r}
	\end{array}
	& \mathbf{H'}
	\end{array}\right]
	\mathbf{P}}_{= \mathbf{H}}  + \mathbf{N},
	\label{eq:LQ_nearSep}
	\end{equation}
	where $\mathbf{W} \in \mathbb{R}^{m\times r}$, %$\ps(\mathbf{W}) \in \mathbb{R}^{m\times \frac{r(r+1)}{2}}$, 
	$\mathbf{I_r}$ is the $r$-by-$r$ identity matrix, $\mathbf{0}_{p \times q}$ the $p$-by-$q$ matrix of zeros,  $\mathbf{P}$ a permutation matrix and $\mathbf{H}' \in \mathbb{R}^{\frac{r(r+1)}{2} \times (n - r)}$ satisfying the two first conditions of (\ref{eq:hypLQEntiere}).
	\label{hyp:NearSep}
\end{assumption}
The aim of this work is to introduce an algorithm which, given a $r$-LQ near separable mixture, recovers the factors $\mathbf{W}$ and $\mathbf{H}$, up to a permutation. To do so, we generalize the SNPA \cite{Gillis2014} by explicitely modeling the bilinear products along the greedy search process.\\

We denote matrices as $\mathbf{A} \in \mathbb{R}^{m\times r}$, a column indexed by $i \in \lbr 1, r \rbr$ as $\mathbf{a}_i$ and a row indexed by $j \in \lbr 1, m \rbr$ as $\mathbf{a}^j$. 
The quantity $|\mathcal{K}|$ is the number of elements in the set $\mathcal{K}$. We define the set $\Delta^r = \{x \in \mathbb{R}^r | x \geq 0, \sum_{i = 1}^{r} x_i \leq 1\}$. 

\section{Proposed \snpab\ algorithm}
The proposed \snpab\ (see Algo. \ref{alg:algo}) is an extension of SNPA \cite{Gillis2014}, which is an algorithm designed for linear near-separable NMF. Similarly to SNPA, \snpab\ is a greedy algorithm. At each iteration, the column of the data matrix $\mathbf{X}$ with the largest $\ell_2$ norm is selected.
\snpab\ and SNPA however differ by their respective projection steps:
\begin{itemize}
	\item SNPA projects each column of $\mathbf{X}$ onto the convex hull formed by the origin and all the columns extracted so far;
	\item In \snpab, we propose to perform the projection of each column of $\mathbf{X}$ on the convex hull formed by the origin, the columns extracted so far \emph{and their second order products}. 
\end{itemize}
Therefore, if two endmembers $\mathbf{w}_i,\ i \in \lbr 1,r \rbr$ and $\mathbf{w}_j,\ j \neq i,\ j \in \lbr 1,r \rbr$ have been extracted during the iterative process of \snpab, the contribution of the quadratic term $\mathbf{w}_i \odot \mathbf{w}_j$ is cancelled. As such, the non-linear part of the mixing is reduced, giving more weight to the linear contribution. Thus the endmembers are expected to be more easily extracted.

\begin{algorithm}[!h]
	\caption{\snpab: Successive Nonnegative Projection Algorithm for Linear Quadratic mixtures.}
	\label{alg:algo}
	\begin{algorithmic}
		\STATE {\bf Input:} A $r$-LQ $r$-near-separable matrix $\mathbf{X}  \in \mathbb{R}^{m\times n}$ satisfying constraints  \eqref{eq:hypLQEntiere}, the number $r$ of endmembers.
	
		\STATE {\bf Initialization:}  $\mathbf{R = X}$, $\mathcal{K} = \{ \}$
		%		\WHILE{$\norm{\mathbf{R}}{1,2} > t$ \AND $l \leq r$}
		\STATE \% \emph{Greedy search}
		\WHILE{$|\mathcal{K}| \leq r$}
		\STATE $p = \argmax{j\in \lbr 1,n \rbr}\norm{\mathbf{r}_j}{2}$
		\STATE $\mathcal{K} = \mathcal{K} \cup \{p\}$
		\FOR{$j \in  \lbr 1,n \rbr$}
		 \STATE $\mathbf{h}_{j} = \argmin{h \in \Delta^{\frac{|\mathcal{K}|(|\mathcal{K}|+1)}{2}}} \norm{\mathbf{x}_j - \ps(\mathbf{X}_\mathcal{K})h}{2}$
		 \STATE $\mathbf{r}_j = \mathbf{x}_j - \ps(\mathbf{X}_\mathcal{K})\mathbf{h}_{j}$
		\ENDFOR
		% ND: vieille version
		%\STATE $\mathbf{R}^j = \mathbf{X}^j - %\ps(\mathbf{X}^\mathcal{K})\mathbf{H}^{*j}$ for all $j$, 
		%\STATE where 
		%\begin{equation}
		%\mathbf{H}^{*j} = \argmin{x \in \Delta^{\frac{|\mathcal{K}|(|\mathcal{K}|+1)}{2}}} f\left(\mathbf{X}^j - \ps(\mathbf{X}^\mathcal{K})x\right)
		%\label{eq:proj}
		%\end{equation}
		\ENDWHILE
		%		\item[]
		%		\STATE {\bf Post-processing:}
		%		\STATE Let $N = \{\}$
		%		\FOR{$k \in \mathcal{K}$}
		%		\STATE $\mathbf{M}$ = $\left[(\mathbf{X}^i)_{i\in \mathcal{K} \setminus \{k\}},(\mathbf{X}^i \odot \mathbf{X}^j)_{i,j\in \mathcal{K}}\right]$
		%		
		%		\STATE $N = N \cup \{\norm{\mathbf{X}^k - \mathbf{M}\mathbf{H}^{*k}}{2}\}$, 
		%		\STATE where $\mathbf{H}^{*k} = \argmin{x \in \Delta^{\frac{r(r+1)}{2} - 1}} f\left(\mathbf{X}^k - \mathbf{M}x\right)$
		%		\ENDFOR
		%		\item[]
		%		\STATE $\mathcal{K}' = \{r \text{ indices in } \mathcal{K} \text{ minimizing } N\}$
		\STATE {\bf Output:} Set $\mathcal{K}$ of indices such that $\mathbf{X}_{\mathcal{K}} \simeq \mathbf{W}$ up to a permutation.
	\end{algorithmic}
\end{algorithm}

\section{Numerical results}
\label{sec:experiments}
The experiments are conducted on a noiseless\footnote{The results are similar when some noise is added, but the study of \snpab\ for various noise level is omitted in this paper due to lack of space.} realistic dataset $\mathbf{X}$ of the form \eqref{eq:LQ_nearSep}. Up to $20$ spectral signatures  are extracted from the USGS database\footnote{https://www.usgs.gov/} to build $\mathbf{W}$ with $m=20$ and $r\in \lbr 2,20 \rbr$. 
The matrix $\mathbf{H}$ has dimension $\frac{r(r+1)}{2} \times 1000$, and the columns of 
$\mathbf{H'}$ in \eqref{eq:LQ_nearSep} are generated randomly using a Dirichlet distribution $\mathcal{D}\left(\alpha,\ldots,\alpha\right)$ with $\alpha=0.5$. The results are averaged over $100$ Monte-Carlo experiments.

\begin{figure}[!h]
\vspace{-0.5cm}
	\centering
	\begin{tikzpicture}
	\node (img1) {\includegraphics[width=2.5in]{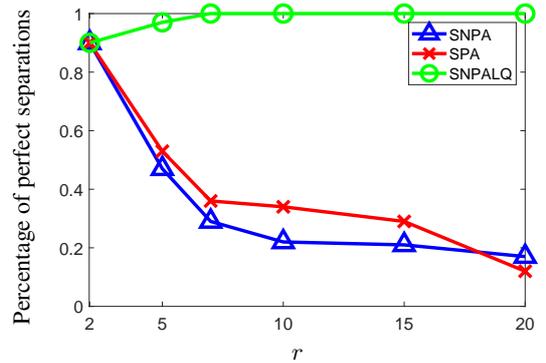}};
	\node[below=of img1, node distance=0cm, yshift=1.2cm] {\small $r$};
	\node[left=of img1, node distance=0cm, rotate=90, anchor=center,yshift=-0.7cm] {\small Percentage of perfect separations};
	\end{tikzpicture}
	\vspace{-0.3cm}
	\caption{Probability of perfect separation as a function of $r$.}
	\label{fig_res_noiselessLQ}
\end{figure}

Given a set of indices $\mathcal{K}$  extracted by and algorithm, the separation quality is assessed using
\begin{equation*}
\theta = 
\min_{i \in \lbr 1,r \rbr} 
\diag 
\left( 
\mathbf{W}^T \mathbf{X}_\mathcal{K} 
\right),
\end{equation*}
where $\diag(\mathbf{A})$ is contains the diagonal elements of the matrix $\mathbf{A}$.
We consider perfect separation is achieved if $\theta > 0.999$.\\
%\subsection{Numerical results}
\label{sec:noiselessExp}
The probability of obtaining a perfect separation using several algorithms is displayed in Fig.~\ref{fig_res_noiselessLQ} as a function of the number $r$ of endmembers.
\snpab\ obtains significantly better results than SNPA~\cite{Gillis2014} or SPA~\cite{Gillis_12_FastandRobust}, especially for large $r$. It achieves a perfect separation in more than $90 \%$ of the experiments. The initial improvement when $r$ increases is linked to the use of a Dirichlet distribution with $\alpha=0.5$ when randomly generating the mixing coefficient matrix $\mathbf{H}'$. When $r$ is small, the datapoints are more spread within the convex hull formed by the origin and the (virtual) endmembers, leading to a higher probability for a virtual endmember to be extracted.\\
SNPA and SPA results deteriorate quickly when $r$ increases. SPA becomes worse than SNPA when $r \simeq m$, which is expected since SNPA has an interest mainly when the endmember matrix $\mathbf{W}$ is either rank-deficient or ill-conditioned~\cite{Gillis2014}.

\section*{Conclusion}
To tackle the problem of linear-quadratic hyperspectral unmixing, we introduced \snpab, an extension of SNPA which explicitly includes the quadratic terms into the projection step. The approach was shown to obtain good results on non-linear realistic datasets. More results, both empirical and theoretical, will be given at the conference, including a study of the proposed algorithm \snpab\  with respect to noise.

%% You can make the bibliography smaller
\bibliographystyle{IEEEtran}
\bibliography{bibliographie}

\end{document}